\documentstyle[pra,aps]{revtex}
\begin{document}
\draft
\twocolumn[\hsize\textwidth\columnwidth\hsize\csname
@twocolumnfalse\endcsname

\title{Hydrodynamic excitations of Bose Condensates
       in anisotropic traps}

\author{Martin Fliesser$^1$, Andr\'as Csord\'as$^2$,
P\'eter Sz\'epfalusy$^{3}$, Robert Graham$^1$}

\address{$^1$Fachbereich Physik,
Universit\"at-Gesamthochschule Essen,
45117 Essen,Germany \\
$^2$Research Group for Statistical Physics of the
Hungarian Academy of Sciences,\\
M\'uzeum krt. 6--8, H-1088 Budapest,
Hungary \\
$^3$Institute for Solid State
Physics, E\"otv\"os University,
M\'uzeum krt. 6--8, H-1088 Budapest,
Hungary, and\\
Research Institute
for Solid State Physics, P.O. Box 49, H--1525 Budapest, Hungary,}

\date{\today}
\maketitle

\begin{abstract}
The collective excitations of Bose condensates in anisotropic axially
symmetric harmonic traps are investigated in the hydrodynamic and Thomas-Fermi
limit. We identify an additional conserved quantity, besides the axial
angular momentum and the total energy, and separate the wave equation in
elliptic coordinates. The solution is thereby reduced to the algebraic
problem of diagonalizing finite dimensional matrices. The classical quasi-particle dynamics in the local density approximation for energies 
of the order of the chemical potential is shown to be chaotic.

\end{abstract}


\vskip2pc]
\narrowtext

The new Bose condensates of alkali atoms in magnetic traps \cite{1,2,3}
offer a unique way to investigate the low-lying collective excitations in
Bose condensates \cite{4,5,6,7,8,8b,9,10,11,12,13,14,15}. Experimentally
collective modes with a given symmetry have been excited by time-dependent
modulations of the trapping potential, and their evolution has been
followed in real time by measurements of the resulting shape oscillations
of the condensates. The measurements performed so far have involved turning off
the trap after a given time \cite{4,5,6}, but in the future they could even
be performed non-destructively by elastic off-resonant light scattering
\cite{7}. Theoretically the collective modes have been analyzed 
by using the Bogoliubov-equations or by linearizing
the time-dependent Gross-Pitaevskii equation around the time-independent
condensate and solving these equations numerically \cite{8,8b} or analytically
in various approximations \cite{9,10,11,12,13,14,15}. Very good agreement
between the numerical and the experimental results has been found.

In a seminal paper Stringari \cite{15} has shown how the coupled wave equations
for the collective excitations are simplified in the hydrodynamic limit to
become a single second-order wave equation for density waves, and he obtained
analytical solutions for all its modes in spherically symmetric harmonic traps,
and, remarkably, also for some of its modes in axially symmetric harmonic
traps. The latter are particularly important, because all experiments have
been performed with traps of this symmetry \cite{4,5,6}.

In the present paper it is our goal to study in more detail by analytical
means the hydrodynamic wave equation in the axially symmetric case. We wish
to find an explanation why at least some analytical solutions have been
possible in this case and intend to use this insight to construct more
solutions in a systematic way.

In principle the collective mode problem looks very different for isotropic
and for axially symmetric traps: In the isotropic case the rotational
symmetry ensures that angular momentum conservation gives two good
quantum numbers, and therefore the wave equation is separable in spherical
coordinates. For axial symmetry, however, only the axial component of angular
momentum remains a good quantum number, besides the energy, and one may
expect that the system, having three degrees of freedom, is not integrable.
In fact, this expectation is born out for collective excitations whose
energies are neither very large nor very small compared to the chemical
potential (see below),
which spoils all hopes to find exact analytical solutions for the modes and
their spectrum on this energy range for axially symmetric traps. Why then
are such solutions possible at energies in the hydrodynamic regime, i.e. for
energies much smaller than the chemical potential?

The answer is provided by the existence, in that regime, of an additional
conserved quantity which we exhibit explicitely below. Its  existence
permits the separation of the hydrodynamic wave equation in elliptical
coordinates. Thereby the task of solving the wave equation can be reduced to
the purely algebraic problem of diagonalizing finite dimensional matrices. We
use this method to obtain the spectrum of the low lying hydrodynamic modes
as a function of the ratio of the axial and the radial trap frequencies.

A convenient starting point of our analysis are the linearized hydrodynamic
equations as derived in \cite{15}. They read
\begin{equation}
\label{eq:1}
\dot{\varphi} = -\frac{4\pi\hbar a}{M}\delta\rho,\quad
\delta\dot{\rho} =-\frac{\hbar}{M}\bbox{\nabla}\cdot\rho_0(\bbox{x})
       \bbox{\nabla}\varphi\nonumber
\end{equation}
Here $\varphi$ is the phase of the macroscopic wave function, $\delta\rho$ the
local perturbation of the number density in the collective mode,
$\rho_0(\bbox{x})$ is the number density of the time-independent condensate
in the Thomas-Fermi approximation \cite{17},
$\rho_0(\bbox{x})=\frac{M}{4\pi\hbar^2a}(\mu-U(\bbox{x}))
\Theta(\mu-U(\bbox{x}))$, where $\mu$ is
the chemical potential, $M$ and $a$
are the mass and the positive $s$-wave scattering length of the atoms. 
For the connection of (\ref{eq:1}) with the Bogoliubov-equations
see \cite{several}.
The trapping potential $U(\bbox{x})$ is assumed to have the form
$U(\bbox{x})=(M/2)\omega_0^2(x^2+y^2)+(M/2)\omega^2_z z^2$.
The surface of the condensate, in the Thomas-Fermi limit is defined by
$U=\mu$. Eqs.~(\ref{eq:1}) are obtained by a gradient expansion and assuming
$\delta\rho$ and $\bbox{\nabla}\varphi$ to be small.
Eliminating $\bbox{\nabla}\varphi$ from both equations and making the ansatz
$\delta\rho(\bbox{x},t)=e^{-i\omega t}\psi(\bbox{x})$ one obtains the
time-independent wave equation \cite{15}
$M\omega^2\psi(\bbox{x})=-\bbox{\nabla}\cdot(\mu-U(\bbox{x}))
  \bbox{\nabla}\psi(\bbox{x})$
which holds inside the condensate. We look for boundary conditions on the
surface of the condensate which make the operator
$\hat{G}=-M^{-1}\bbox{\nabla}\cdot(\mu-U(\bbox{x}))\bbox{\nabla}$ Hermitian, so that for two eigenfunctions
$\psi$, $\tilde{\psi}$ with eigenvalues $\omega^2$, $\tilde{\omega}^2$ we have
orthogonality according to
\begin{eqnarray}
(\omega^2 &-& \tilde{\omega}^2)\int_Vd^3x\psi^*\tilde{\psi}=
\int_Vd^3x[\psi^*\hat{G}\tilde{\psi}-\tilde{\psi}\hat{G}^*\psi^*]\nonumber\\
          &=& -M^{-1}\int_{\partial V}df(\mu-U)
 (\psi^*\partial\tilde{\psi}/\partial n-\tilde{\psi}
  \partial\psi^*/\partial n)=0 \,\nonumber.
\end{eqnarray}
Here $V$ and $\partial V$ denote the volume of the condensate and its
surface, respectively. Because $\mu-U$ vanishes on the boundary it is enough
to require that $\psi$ and its normal derivative $\partial\psi/\partial n$
remain bounded there.
In the following it will be useful to measure lengths in units of the radial
Thomas-Fermi radius $r_0=(2\mu/M\omega_0^2)^{1/2}$, which brings the wave equation
into the dimensionless form $\omega^2\psi=\hat{G}\psi$ with
\begin{equation}
\label{eq:5}
 \hat{G}=-(\omega^2_z/2)\bbox{\nabla}\cdot
  \left((1-\epsilon^2)(1-\rho^2)-z^2\right)
  \bbox{\nabla}
\end{equation}
where we use cylindrical coordinates $\rho=\sqrt{x^2+y^2},z$ and the azimuthal
angle $\phi$ and define $\epsilon^2=1-\omega_0^2/\omega_z^2$, which is
positive for $\omega_0<\omega_z$. The operator $\hat{G}$ commutes, of course,
with the axial angular momentum operator
$\hat{L}_z=-i\frac{\partial}{\partial\phi}$. However, there is an additional
nontrivial operator
\begin{equation}
\label{eq:6}
 \hat{B}=-\bbox{\nabla}^2+(\bbox{x}\cdot\bbox{\nabla})^2+3\bbox{x}\cdot
  \bbox{\nabla}+
 \epsilon^2\partial^2/\partial z^2
\end{equation}
which commutes with $\hat{L}_z$ and with $\hat{G}$, as one may check by
direct calculation of the commutators. In the isotropic case $\epsilon=0$,
$\hat{B}$ may be expressed by the square of the angular momentum
$\hat{L}^2=\hat{L}_x^2+\hat{L}_y^2+\hat{L}_z^2$ and $\hat{G}$ via
%\begin{equation}
%\label{eq:7}
 $\hat{B}=\hat{G}/\omega_0^2+\hat{L}^2$.
%\end{equation} 
Because
of the existence of the three commuting operators $\hat{G}$, $L_z$, and
$\hat{B}$ in the system with the three degrees of
freedom $\rho$, $z$, $\phi$ it is now manifest that the system is
integrable, i.e. the eigenvalues of $\hat{G}$ can be labelled by the
quantum numbers of $L_z$ and $\hat{B}$. To see this explicitely we now look
for variables in which the wave equation separates and introduce
cylindrical elliptical coordinates $\xi$, $\eta$, which in their oblate
spheroidal form \cite{18} are defined by
$ \rho=\sigma\sqrt{(\xi^2+1)(1-\eta^2)}\,,\quad z=\sigma\xi\eta$.
These coordinates are orthogonal. Surfaces of $\xi=$ const are confocal
ellipsoids with foci at $z=0$, $\rho=\sigma$. Surfaces with constant $\eta$
are confocal hyperboloids with the same foci. For $\omega_z\ge\omega_0$,
i.e. $0\le\epsilon^2\le 1$, the foci at $z=0$ $\rho=\sigma$ are made to
coincide with the foci of the ellipsoidal Thomas-Fermi surface if we choose
$\sigma=\epsilon$. The Thomas-Fermi surface is given by
$\xi_{TF}=(1/\epsilon^2-1)^{1/2}$.
Then the interior of the condensate is described by $\xi$ in the range
$[0,\xi_{TF}]$ and $\eta$
in the range $[-1,1]$. 

For $\omega_z\le\omega_0$ the parameter
$\epsilon^2$ is no longer usefull as it becomes negative. Instead one can
define $\epsilon^{'2}=1-\omega_z^2/\omega_0^2=-\epsilon^2/(1-\epsilon^2)$
which lies in the range $0\le\epsilon^{'2}<1$. The foci of the Thomas-Fermi
ellipse now lie at $z=\pm\sigma'$ where $\sigma'=\epsilon'/\sqrt{1-\epsilon^{'2}}$. Therefore we now  
need the prolate spheroidal form of elliptical coordinates
\cite{18} with foci at $z=\pm\sigma'$, $\rho=0$.
These coordinates are defined by
$\rho=\sigma'\sqrt{(\xi^2-1)(1-\eta^2)}\,,
\quad z=\sigma'\xi\eta$
where inside the condensate $\xi$ now has the range $[1,1/\epsilon']$ while
$\eta$ has the same range as before. The treatments in the two cases are equivalent via the transformation connecting $\epsilon$ and $\epsilon'$. In the following we shall present the equations for the
case $\omega_z>\omega_0$. The final formulas for $\omega^2$ apply for
$\omega_z\ge\omega_0$ and $\omega_z\le \omega_0$. After the change of coordinates with  
$\sigma=\epsilon$ the operator $\hat{G}$ takes the
form
\begin{eqnarray}
  \label{eq:11}
 & &\hat{G}=     {\omega_z^2 \over 2 \epsilon^2}\frac{1}{\xi^2 + \eta^2}
      \\
 & &
 \biggl[(1-\epsilon^2(1-\eta^2))\frac{\partial}{\partial \xi}
       (1-\epsilon^2(\xi^2+1))(\xi^2+1)\frac{\partial}{\partial \xi}\biggl.
     \nonumber\\
 \biggl.&+& (1-\epsilon^2(\xi^2+1))\frac{\partial}{\partial \eta}
       (1-\epsilon^2(1-\eta^2))(1-\eta^2)\frac{\partial}{\partial \eta}\biggl.
     \nonumber\\
 \biggl.&+&\frac{(1-\epsilon^2(\xi^2+1))(1-\epsilon^2(1-\eta^2))
 (\xi^2+\eta^2)}{(1-\eta^2)(1+\xi^2)}
    (\frac{\partial}{\partial \phi})^2 \biggr]\nonumber
\end{eqnarray}
which is now separable. The $\phi$-dependence of its eigenfunctions is taken
care of by factors $e^{im\phi}$ with the integer azimuthal quantum number $m$.
Separating the operator in $\xi$ and $\eta$ by making the ansatz
$\Psi_\xi(\xi)\Psi_\eta(\eta)e^{im\phi}$ for its eigenfunctions we obtain
two equations, one for $\Psi_{\eta}$
\begin{eqnarray}
  \left[\frac{d}{d \eta}(1- \eta^2)\frac{d}{d \eta}
  -\frac{m^2}{1- \eta^2}
  +{2\epsilon^2 (1-\eta^2)\eta \over 1-\epsilon^2 (1-\eta^2)} \frac{d}{d \eta}
  \right]\ & &\Psi_{\eta} \nonumber\\
  -{2 \omega^2 / \omega_1^2 \over 1-\epsilon^2 (1-\eta^2)}\Psi_{\eta}
  = -\beta & &\Psi_{\eta}\, .
\label{eq:12}
\end{eqnarray}
the other for $\Psi_{\xi}$. It turns out that both equations are identical if  
in the equation for $\Psi_{\eta}$ we substitute $i\xi$
for $\eta$, i.e. $\Psi_\xi(\xi)\equiv \Psi_\eta(i\xi)$. The solution for one
coordinate is the analytic continuation of the solution of the other from the
real to the imaginary axis. It is easy to check that the separation constant  
$\beta$ is just the eigenvalue
of the operator $\hat{B}$ for the eigenfunction  
$\Psi_{\xi}(\xi)\Psi_{\eta}(\eta)\exp(im\phi)$. To do this one needs
to express $\hat{B}$ also in the elliptic coordinates:
\begin{eqnarray}
\label{eq:13}
\hat{B}&=& \frac{1}{\epsilon^2(\xi^2+\eta^2)}\Bigg\{
  -\frac{\partial}{\partial \xi}(1-\epsilon^2(\xi^2+1))(\xi^2+1)\frac{\partial}{\partial \xi}\nonumber\\
  &-&\frac{\partial}{\partial\eta}(1-\epsilon^2(1-\eta^2))(1-\eta^2)\frac{\partial}{\partial\eta}\nonumber\\
   &+&\frac{\xi^2+\eta^2}{(\xi^2+1)(1-\eta^2)}
    \frac{\partial^2}{\partial\phi^2}\Bigg\}\,.
\end{eqnarray}

The equations (\ref{eq:12}) contain $m$ only quadratically. Therefore the
energy levels are the same for $\pm m$. 
Expanding $\Psi_\eta$ in terms of associated
Legendre functions 
\begin{equation}
\Psi_\eta=\sum_{\ell=|m|}^\infty a_\ell{P_\ell}^{|m|}(\eta)
\label{expansion} 
\end{equation}
we obtain from eq.~(\ref{eq:12})
a second order recursion relation for the coefficients $a_\ell$
relating only even or only odd indices $\ell$. The eigenstates therefore have
even and odd parity. The recursion relation itself is straightforward to
obtain but lengthy and will not be written out here. 
The condition that (\ref{expansion}) terminates at 
 $\ell_{max}=(|m|+n)$
quantizes the eigenvalue $\beta$ 
\begin{equation}
\label{eq:15}
 \beta=(n+|m|)(n+|m|+3)\,,
\end{equation}
which means that $\Psi_\eta$ becomes $(1-\eta^2)^{|m/2|}$ times a
polinomial of order $n$.

In the {\it isotropic} case $\epsilon=0$ the operator $\hat{B}$ can be
diagonalized in spherical coordinates and its spectrum then is found as
$\beta=(2n_r+\ell)(2n_r+\ell+3)$ with radial quantum number $n_r$ and angular
quantum number $\ell$. Together with the connection of $\hat{G}$
and $\hat{B}$ for isotropic traps this gives the result
of \cite{15} for the spectrum in the isotropic case. 
%It also shows the meaning
%of the new quantum number $n$ in that case, where it becomes
%$n=2n_r+\ell-|m|$ and is clearly associated with the motion
%transverse to the $z$-axis. 

The eigenvalue condition for $\omega^2$
takes the form of the
characteristic equation of a tridiagonal  matrix of dimension
$N=1 +\mbox{\rm int}\,[n/2]$, which can be symmetrized by a suitable
similarity transformation with a given diagonal matrix. For fixed
numbers $n$,
$|m|$ we have $N$ different solutions for $\omega^2$, which we label by our
third quantum number $j=0,\dots\mbox{\rm int}\,[n/2]$. In the {\it isotropic}
case $\epsilon=0$ the quantum number $j$ can be expressed as
$j=\mbox{\rm int}[(\ell-|m|)/2]=\mbox{\rm int}[n/2]-n_r$
as one finds by expressing the isotropic spectrum of \cite{15} in terms of
the new quantum numbers $n$, $j\le\mbox{\rm int}\,[n/2]$. 
%$j$ is asociated
%with the axial motion along the $z$-axis.

Calculating the first levels $\omega(n,j,m)$ we get with  
$\lambda=(\omega_0/\omega_z)^2$
\begin{eqnarray}
\label{eq:16}
  &&\omega^2(0,0,m)=\omega_0^2 |m|
  \nonumber\\
 && \omega^2(1,0,m)=\omega_z^2 +\omega_0^2|m|
  \nonumber\\
  &&\omega^2(2,j(=0,1),m)=\omega_z^2 (\frac{3}{2}+2(|m|+1)\lambda
  \nonumber\\
                 &&\qquad -\frac{(-1)^j}{2} (9-4(|m|+4)\lambda
                                + 4(|m|+2)^2\lambda^2)^{1/2})
  \nonumber\\
  &&\omega^2(3,j(=0,1),m)=\omega_z^2 (\frac{7}{2}+2(|m|+1)\lambda
  \nonumber\\
                 && \qquad -\frac{(-1)^j}{2}(25+4(|m|-4)\lambda
                                + 4(|m|+2)^2\lambda^2)^{1/2})\, .
\end{eqnarray}
In the limit $\lambda^{-1} \to 0$, which is relevant for the experiments reported in \cite{5}, the mode frequencies for {\it arbitrary} integer $n\ge 2j$ not to large can be expanded in the small parameter $\lambda^{-1}$ and are found as
\begin{equation}
\label{eq:as}
 \omega^2(n,j,m)=\omega_0^{2}(|m|+2j(j+|m|+1)+O(\lambda^{-1}))
\end{equation}
Remarkably, the leading order term for $|m|,j$ not both vanishing is independent of $n$, i.e. the levels consist in this case of bands of closely spaced levels split only by small frequencies of order $\omega_z^2/\omega_0$.

Eq.(\ref{eq:16}) contains, as special cases, the particular solutions previously obtained
by Stringari \cite{15} (the $n=0,1; j=0,1$ modes for all $|m|$ and the
two $n=2; j=0,1; m=0$ modes). 
%The results (\ref{eq:16}) can be compared
%with the result for the free trap
%$\omega=\omega_0\left(2\,\mbox{\rm int}\,[n/2]
%+|m|-2j)+\omega_z(n-2\,\mbox{\rm int}\,\left[\frac{n}{2}\right]+2j\right)$
%expressed in the same quantum numbers.

After our calculation had been
completed and while this paper was being prepared a preprint became available
\cite{19}, in which the mode frequencies
(\ref{eq:16}) where also found by solving the wave equation directly via a
polynomial ansatz, which can be shown to be equivalent to ours. 
This method works because, as we have shown above, no
special boundary conditions except regularity need to be imposed on the wave
function at the surface of the condensate. However, the deeper {\it reason}
for the solvability of the equation, i.e. the additional conservation law,  
has not been identified in \cite{19}.

Because of the necessity to diagonalize $N$-dimensional matrices the
analytical determination of the $\omega^2$ by the present method for $\lambda$ neither very small nor very large is possible
up to $N=4$, even though the formulas for $N=3$ and $N=4$ are too
cumbersome to be of much practical value. However, it is straightforward to
diagonalize the matrices numerically and to prepare a list of the numerical
values of the eigenvalues for a given ratio $\lambda$,
e.g. for the experimental value $\lambda=1/8$. In table I we give such a list
for the 43 lowest lying eigenvalues with $|m|\le 2$ in units of $\omega_0$ for
$\lambda=1/8$, together with their quantum numbers $m$, $n$, $j$.
A numerical evaluation of the energy levels of the collective modes as a function of the
scaling variable \cite{15} $Na/d_0$ for $\lambda=1/8$ was recently reported in
\cite{8b}. Here $N$ is the number of atoms and $d_0=\sqrt{\hbar/2M\omega_0}$
is a measure of the size of the ground state in the trap. The present results
apply for $Na/d_0\to\infty$. Comparing our results with those in \cite{8b}
we find that the rate of convergence to the asymptotic case $Na/d_0\to\infty$
depends on the quantum number $n$ and is much slower, e.g. for $n=6$ than for $n=2$.
For $m=0$ we find 4 additional levels (with quantum numbers $n=5,7,8,9$
and $j=0$) below the highest m=0 level considered in \cite{8b}, where the motion of
the levels was tracked from the free particle case to the strongly interacting
case.
The explanation can be that levels
% with large $n$ and $m=0=j$, 
having rather high frequency in the free trap 
%($\omega=4\omega_0+\omega_z$ for the missing
%$n=5$ mode) but 
can move down into the considered low-energy regime as the
parameter $Na/d_0$ is increased. 
%%%%%%%%%%%%%%%%%%%%%%%%%%%%%%%%%%%%%%%%%%%
Following the levels one can expect a number of avoided level
crossings due to the non-integrability of the system at intermediate
energies. We demonstrate the chaoticity of the corresponding
classical system whose Hamiltonian
\begin{eqnarray}
H(\bbox{p},\bbox{x})&=&\sqrt{\epsilon_{HF}^2(\bbox{p},\bbox{x})-
K^2(\bbox{x})}, \nonumber\\
\epsilon_{HF}(\bbox{p},\bbox{x})&=&{\bbox{p}^2 \over 2M}+
U(\bbox{x})-\mu+2 K(\bbox{x}) \nonumber\\
K(\bbox{x})&=&(\mu-U(\bbox{x}))\Theta(\mu-U(\bbox{x}))
\end{eqnarray}
is the Bogoliubov quasiparticle energy in local density approximation \cite{zus}
(Fig.1)
%%%%%%%%%%%%%%%%%%%%%%%%%%%%%%%%%%%%%%%%%%%%

In conclusion we have demonstrated that the quasiparticle dynamics in axially symmetric traps is chaotic for energies comparable to the chemical potential, but, for energies much smaller than $\mu$ we found a third conserved variable $\hat{B}$ commuting
with the wave operator $\hat{G}$ and axial angular momentum $\hat{L}_z$.
As a consequence the wave equation in this regime is proven to be integrable which
explains why some solutions have already been found in the literature. We
have separated the wave equations in elliptical coordinates, 
reduced the eigenvalue problem to the diagonalization of finite dimensional
matrices, and exhibited three integer quantum numbers $m$, $n$, $j$ in the
ranges $-\infty\le m\le +\infty$, $0\le n<\infty$, $0\le j\le \mbox{\rm int}\,
[n/2]$
labelling all modes and their frequencies for the experimentally realized case
$\omega^2_z/\omega^2_0=8$.
As a final remark we mention that we have found even the fully anisotropic
case, where the axial angular momentum is no longer a good quantum number, to
be integrable \cite{20}. {\it Two} new conserved quantities can be identified in
that case which reduce to $\hat{L}_z$ and $\hat{B}$ in the axially symmetric
limit.

\section*{Acknowledgements}

One of us (R.G.) acknowledges discussions with J. Ruostekoski. This work has  
been supported by the
project
of the Hungarian Academy of Sciences and the Deutsche
Forschungsgemeinschaft under grant No. 95.
M.F.\ gratefully acknowledges support
by the German-Hungarian Scientific and Technological Cooperation
under Project 62.
R.G. and M. F. wishes to acknowledge support by the Deutsche
Forschungsgemeinschaft through the Sonderforschungsbereich 237
``Unordnung und gro{\ss}e Fluktuationen''.
Two of us (A.Cs,P.Sz) would like to acknowledge support by
the Hungarian Academy of Sciences under grant No. AKP 96-12/12.
The work has been partially supported by the Hungarian
National Scientific Research Foundation under grant
Nos. OTKA T017493 and F020094.

\begin{figure}
\caption{Poincar\'e sections of the classical dynamics in cylindrical coordinates $\rho,z,\phi$ after the elimination of the conserved axial 
angular momentum and the azimuthal angle $\phi$
for energy $E/\mu=1$. 
The cut is taken at $z=0$ and diplayed in the scaled variables
$\rho,p_{\rho}$ measured in units of $\sqrt{2\mu/m\omega_0^2}$ and $\sqrt{2m\mu}$, respectively.
The anisotropy is chosen as $\omega_z/\omega_0 = \protect\sqrt{8}$,
the angular momentum was fixed as $\omega_0 L_z/E=0.2$.
\label{poinc.cuts}}
\end{figure}
\begin{table}
\begin{tabular}{rrlll}
n  & j & $m=0$ & $m=1$ & $m=2$ \\  \hline

   0  &  0   &   0.00000000 &     1.00000000 &     1.41421356\\
   2  &  0   &   1.79712837 &     2.31729458 &     2.72341601\\
   1  &  0   &   2.82842712 &     3.00000000 &     3.16227766 \\
   4  &  0   &   2.91193010 &     3.36537641 &     3.73131058\\
   3  &  0   &   3.27302589 &     3.51832107 &     3.74165739\\
   5  &  0   &   3.82657318 &     4.09894174 &     4.34324821\\
   6  &  0   &   3.84099626 &     4.21532329 &     4.52279302\\
   7  &  0   &   4.40748368 &     4.67921086 &     4.92221757\\
   8  &  0   &   4.59081095 &     4.89814704 &     5.16010207\\
   9  &  0   &   4.97105460 &     5.22935896 &     5.46144995\\
   2  &  1   &   4.97697997 &     5.16044047 &     5.34630763\\
   10 &  0   &   5.20648119 &     5.47081354 &     5.70417681\\
    4 &  1   &   5.45030761 &     5.74456265 &    (6.04142431)\\
   11 &  0   &   5.49925114 &     5.74053518 &     5.95925695\\
   12 &  0   &   5.73868980 &     5.97616320 &    (6.19077758)\\
\end{tabular}
\caption{The 43 lowest levels labeled by the quantum numbers $|m|\le 2,n,j$
for $(\omega_z/\omega_0)^2 = 8$}
\label{tab.1}
\end{table}

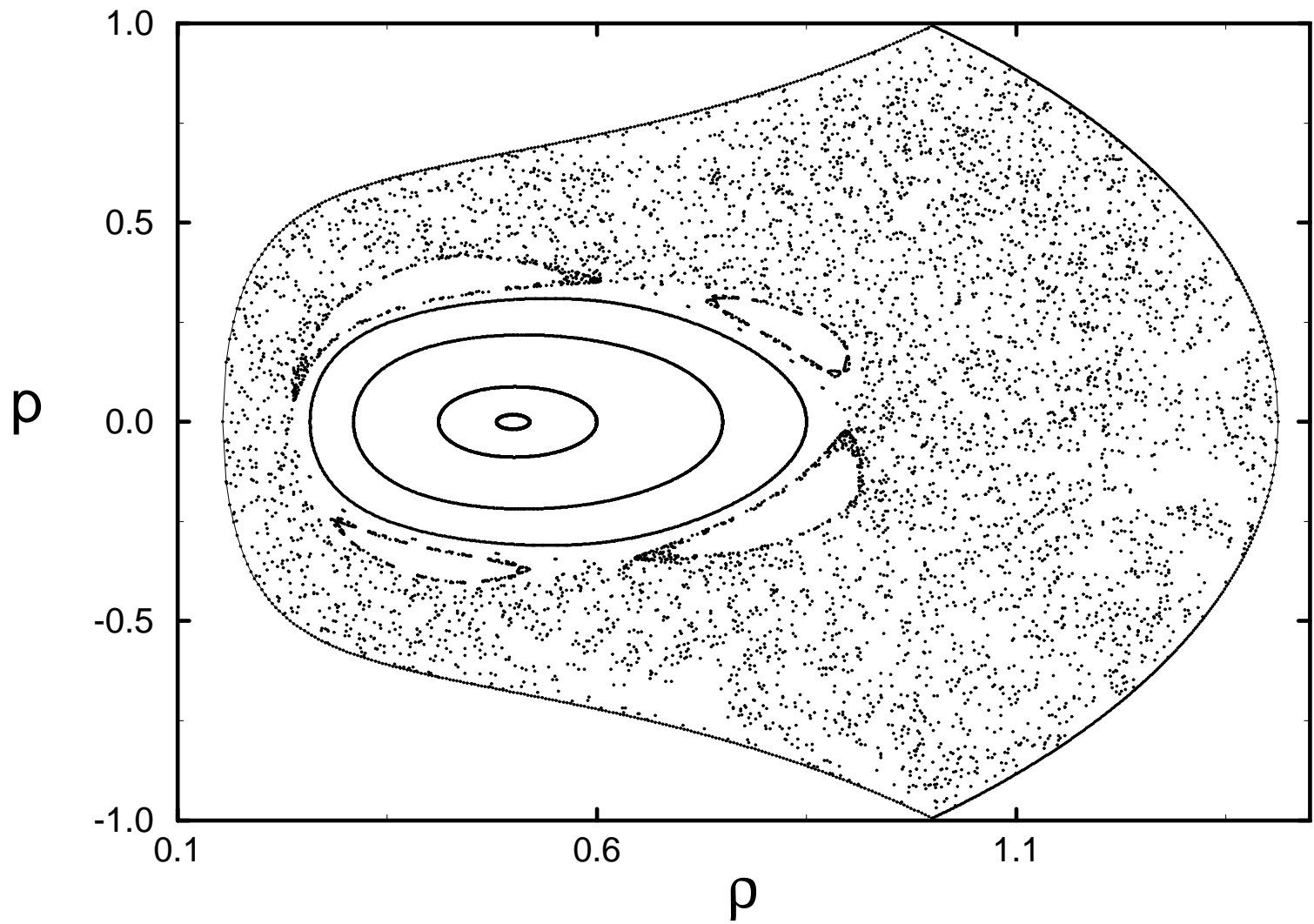
\end{document}